\documentclass[aps,pra,twocolumn,draft]{revtex4}

\usepackage[english]{babel}
\usepackage[dvips]{graphicx}
\usepackage{enumerate,amsthm,amsmath,amssymb,color,graphicx,bbm,float}
\usepackage{natbib}

\newcommand{\ket}[1]{\left| #1 \right\rangle}

\newcommand{\proj}[1]{| #1 \rangle \! \langle #1 |}

\newtheorem{theorem}{Theorem}

\begin{document}
\title{A quantum de Finetti theorem in phase space representation}

\author{Anthony Leverrier} \affiliation{Institut Telecom / Telecom ParisTech,   CNRS LTCI, 46, rue Barrault, 75634 Paris Cedex 13, France}
 
\author{Nicolas J. Cerf} \affiliation{Quantum Information and Communication,   Ecole Polytechnique, CP 165/59, Universit\'e Libre de Bruxelles, 50   av. F. D. Roosevelt, B-1050 Brussels, Belgium} \affiliation{M.I.T. -   Research Laboratory of Electronics, Cambridge MA 02139, USA}

\date{\today}

\begin{abstract}
  The quantum versions of de Finetti's theorem derived so far express the   convergence of $n$-partite symmetric states, i.e., states that are   invariant under permutations of their $n$ parties, towards probabilistic   mixtures of independent and identically distributed (i.i.d.) states of the   form $\sigma^{\otimes n}$. Unfortunately, these theorems only hold in   finite-dimensional Hilbert spaces, and their direct generalization to   infinite-dimensional Hilbert spaces is known to fail.  Here, we address   this problem by considering invariance under orthogonal transformations in   {\em phase space} instead of permutations in {\em state space}, which leads   to a new type of quantum de Finetti's theorem that is particularly relevant   to continuous-variable systems.  Specifically, an $n$-mode bosonic state   that is invariant with respect to this continuous symmetry in phase space   is proven to converge towards a probabilistic mixture of i.i.d. Gaussian   states (actually, $n$ identical thermal states).
\end{abstract}


\maketitle

\section{Introduction}

There has been a renewed interest in de Finetti's theorem \cite{df37,MC93} over the recent years, especially in the context of quantum information theory (see, e.g., \cite{R07}).  In a classical setting, de Finetti's theorem addresses the statistics of large composite systems obeying some fundamental symmetry (e.g., invariance under permutations of its parts), stating that its parts can be well approximated by identical independent subsystems. In the language of probability theory, a permutation-invariant joint probability distribution of $n$ random variables is shown to approach a probabilistic mixture of {\em independent and identically distributed} (i.i.d.) variables. In a quantum setting, the theorem makes the connexion between two types of $n$-mode states in $\mathcal{H}^{\otimes n}$: symmetric states, i.e., states that are invariant under permutations of their subsystems ($\rho$ such that $\rho = \pi \rho \pi^{\dagger}$ for any permutation $\pi \in \mathcal{S}_n$), and mixtures of i.i.d. states of the form $ \sigma^{\otimes n}$ for some state $\sigma \in \mathcal{H}$.  Whereas an i.i.d. state is obviously symmetric, the converse is not true in general. This situation is rather frustrating as the symmetry of a state is often known, or can be easily enforced by application of a random permutation of the subsystems, while it rather is the i.i.d. property that one wishes to have as it considerably simplifies the analysis (an i.i.d. state is fully described in $\mathcal{H}$ instead of $\mathcal{H}^{\otimes n}$). According to the quantum de Finetti's theorem \cite{HM76,CFS02}, a symmetric state becomes increasingly close to a mixture of i.i.d. states as one traces out more of its parts. Attempts at characterizing the speed of convergence towards an i.i.d. state are more recent, both in the classical case \cite{DF80} and quantum case \cite{KR05,CKMR07}: the trace distance between the partial trace over $(n-k)$ parties of an $n$-partite symmetric state and a mixture of $k$-partite i.i.d. states is bounded from above by $2 d^2 k/n$, where $d$ is the dimension of the Hilbert space.

Interestingly, a striking difference with the classical case is that the trace distance in the quantum case necessarily depends on the dimension of the Hilbert space. In particular, this rules out the possiblity of a direct generalization of the theorem to an infinite-dimensional Hilbert space. This was proven in Ref.~\cite{CKMR07}, where a counter-example was exhibited: the $n$-dimensional generalization of the singlet state $1/\sqrt{n!} \; \sum_{\pi} \mathrm{sign}(\pi) \; \pi(|0\rangle \otimes |1\rangle \otimes \cdots \otimes |n-1\rangle)$ is symmetric but any bipartite part, being a mixture of singlet states, cannot be approximated by a mixture of i.i.d. states. Even if a general quantum de Finetti's theorem does not hold in infinite dimension, it is still possible to establish interesting versions of the theorem by restricting the set of states considered. For instance, such results can be obtained for coherent cat states \cite{DOS07} and Gaussian states \cite{KW09}.

In this paper, we follow a rather different approach by considering a symmetry group different from the permutations over the $n$ subsystems of a state in $\mathcal{H}^{\otimes n}$. We investigate the properties of {\em   orthogonally-invariant} states $\rho$, i.e., states that are invariant under the action of any $n$-mode Gaussian unitary operator corresponding to a real symplectic orthogonal transformation in the $2n$-dimensional phase space of $\rho$. In \cite{LKGC09}, we had touched this question in the asymptotic limit $n\to \infty$, and exhibited the connection between orthogonally invariant states and (probabilistic mixtures of) i.i.d. thermal states. Here, we prove a finite version of this result, which leads to a genuine quantum continuous-variable de Finetti's theorem in phase space representation.

The outline of the paper is as follows. In Section \ref{ri-states}, we introduce the concept of orthogonally invariant quantum states, and give an alternative characterization of these states in the Fock state representation. Then, in Section \ref{theorem}, we prove a quantum de Finetti theorem for orthogonally invariant $n$-mode states, which bounds the convergence speed towards i.i.d. thermal states for finite $n$. Finally, in Section \ref{conclusion}, we discuss the perspectives of this continuous-variable quantum de Finetti theorem and draw conclusions.

\section{Orthogonally invariant states}
\label{ri-states}

The state $\rho$ of an $n$-mode bosonic quantum system can be completely characterized by its Wigner function in the $2n-$dimensional phase space parametrized by the quadratures $x_1,p_1,\ldots,x_n,p_n$, namely
\begin{align}
  &W(x_1,p_1, \cdots ,x_n,p_n) \nonumber  \\
  &= \frac{1}{\pi^n} \int_{-\infty}^{\infty} dy_1 \cdots dy_n \; e^{i(p_1 y_1     + \cdots +p_n y_n)}
  \nonumber  \\
  &\times \langle x_1-y_1,\cdots ,x_n-y_n | \rho | x_1+y_1,\cdots ,x_n+y_n   \rangle.
\end{align}
The Wigner function is well known to be a quasi-probability distribution, not a genuine probability distribution as it can take negative values. However, by integrating it over one quadrature ($x$ or $p$) for each mode, one obtains the $n$-variate probability distribution characterizing the outcomes of the $n$ homodyne measurements (one performed on each mode).  For instance $\iiint dp_1 dx_2 dp_3 \; W(x_1,p_1,x_2,p_2,x_3,p_3)$ is the joint probability distribution for the outcomes of the homodyne measurements of quadratures $x_1$, $p_2$, and $x_3$.

One is of course not restricted to measuring quadratures $x_k$ or $p_k$, but can also measure rotated quadratures with any angle $\theta_k$ in phase space. Thus, from a Wigner function, one can always construct a genuine probability distribution $p(r_1, \cdots, r_n)$, where $r_k$ corresponds to a particular rotated quadrature of the $k^{\mathrm{th}}$ mode. In addition, one can also mix several modes with the help of a passive linear interferometer before performing the homodyne measurements, which means that the variables $r_k$ become (normalized) linear combinations of the quadratures $x_1,p_1,\ldots,x_n,p_n$. In summary, starting with an arbitrary Wigner function, one can always construct a family of $n$-variate probability distributions $p(r_1,\cdots, r_n)$ using the following procedure: first, one process the $n$ modes through a passive linear interferometer (a network of beamsplitters and phase shifters), and then one measures one fixed quadrature for each output mode.

Let us now consider possible symmetries of the joint probability distribution characterizing the $n$ random variables $r_k$.  A first symmetry, which is standard in the context of de Finetti's theorem, is the invariance under permutations of the variables. This means that $p(r_1, \cdots, r_n) = p(r_{\pi(1)}, \cdots, r_{\pi(n)})$ for any permutation $\pi \in \mathcal{S}_n$, which denotes the group of permutations on $\{ 1,\ldots,n\}$. Another symmetry, which has not been explored so far in the quantum context, emerges naturally if one considers the real-valued random vector ${\bf r} = (r_1,\cdots,r_n) \in \mathbb{R}^n$. Note that the previous permutation symmetry simply means that the distribution probability is not affected by reordering the coordinates.  As we work in $\mathbb{R}^n$, however, it seems more appropriate to substitute a discrete symmetry group such as $\mathcal{S}_n$ with a continuous symmetry group.  A natural choice in this respect is the orthogonal group $O(n)$, that is, the group of orthogonal transformations (or isometries) acting on vector ${\bf r}$. Note that applying an orthogonal transformation on ${\bf r}$ precisely corresponds to inserting an $n$-mode passive linear interferometer before performing the $n$ homodyne measurements.

In classical probability theory, distributions that are invariant under orthogonal transformations are referred to as {\em orthogonally invariant} distributions.  It has long been known that such probability distributions tend to mixtures of i.i.d. Gaussian distributions in the limit $n\to \infty$, or, more formally, that the first $k$ coordinates of a random point that is uniformly distributed on the $n$-dimensional sphere are asymptotically normal.  (An historical perspective of this property, going back to Poincaré, Borel, and Maxwell, can be found in Ref.~\cite{DF87}, where the authors also derive a sharp bound for the theorem).  In what follows, we consider the natural quantum counterpart of these distributions, namely $n$-mode states $\rho$ for which the probability distribution $p(r_1,\cdots, r_n)$ that results from measuring $n$ quadratures of $\rho$ is unaffected by an $n$-mode passive interferometer preceding the measurement. This is equivalent to the condition that the state $\rho$ is itself invariant under passive symplectic transformations, or, more physically, that $\rho$ remains unchanged after being processed via any $n$-mode passive linear interferometer. In what follows, we will refer to these states as {\em orthogonally invariant} in phase space.

This orthogonal invariance in phase space clearly encompasses the permutation invariance in state space since permuting the coordinates in phase space is just a special case of an orthogonal transformation. Since we are considering a continuous instead of a discrete symmetry group, this invariance in phase space might appear quite restrictive, and we may question whether there exist interesting orthogonally invariant states. This is fortunately the case as, for example, any multimode thermal state is orthogonally invariant. This can be readily checked by considering its Wigner function which is given by a 2$n$-partite Gaussian distribution with variance $\sigma^2$,
\begin{align}
  W_{\mathrm{th}}& (x_1,p_1,\cdots,x_n,p_n) = \frac{1}{(2\pi\sigma^2)^{n/2}}
  \nonumber  \\
  &\times e^{-\left(x_1^2+p_1^2+\cdots + x_n^2+p_n^2\right)/2\sigma^2}
\end{align}
which is clearly invariant under orthogonal transformations of the coordinates. Note that such a multimode thermal state is nothing but a product of identical thermal states, which, in fact, plays the same role for the invariance under orthogonal transformations as i.i.d. states for the usual invariance under permutations. Another class of orthogonally invariant states is, for example, the multimode extension of Fock states that we will consider in the following.

Let us now give two alternative characterizations of the set of orthogonally invariant states.  The most natural one relies on phase space representation, since this is how the symmetry is expressed.  In order to be invariant under orthogonal transformations in phase space, these states must simply have a Wigner function that only depends on one single parameter, namely the modulus $||{\bf r}|| = (x_1^2+p_1^2+ \cdots + x_n^2+p_n^2)^{1/2}$.  The characterization of this set of states in the Fock state representation is slightly more involved.  We note that this set is convex as any mixture of orthogonally invariant states remains invariant under orthogonal transformations. It is, therefore, completely characterized by its extremal points, which can be shown to be the states
\begin{equation}
  \sigma_p^{n} = \frac{1}{a_p^n} \sum_{\stackrel{p_1 \cdots
      p_n}{\;\;\mathrm {s.t.}\; \sum_i p_i = p } } \proj{p_1 \cdots p_n}
  \label{defin}
\end{equation}
with $a_p^n = {n+p-1 \choose n-1}$.  These extremal states are the multimode generalization of number states $\ket{p}$, that is, they correspond to the (normalized) projectors onto the various eigenspaces of the total number operator $\hat{n} = \hat{n}_1 + \ldots +\hat{n}_n$. For instance, $\sigma_p^{n}$, which is proportional to the projector onto the eigenspace of $\hat{n}$ with eigenvalue $p$, physically corresponds to a state with $p$ photons distributed over $n$ modes. The normalization factor $a_p^n$ simply coincides with the number of ways of distributing $p$ photons over $n$ modes. These extremal states $\sigma_p^{n}$ form a discrete infinite set of mixed states parametrized by $p$ (or pure states for $n=1$ as $\sigma_p^{1}=\proj{p}$).  Importantly, any pure eigenstate chosen in the eigenspace corresponding to a given total photon number $p$ is generally not orthogonally invariant; only the uniform mixture of them fulfills this invariance (Schur's lemma), which is why the extremal states $\sigma_p^{n}$ are mixed for $n>1$.

\section{A quantum de Finetti theorem for orthogonally invariant states}
\label{theorem}

As mentioned above, a classical de Finetti's theorem exists for classical orthogonally invariant probability distributions. The theorem states that, in the limit of infinite sequences $X_1,\cdots,X_n$ with $n\to\infty$, the first $k$ variables are exactly mixtures of i.i.d. Gaussian distributions.
This result only holds approximately for finite sequences \cite{DF87}: if the distribution of $X_1,\cdots,X_n$ is invariant under orthogonal transformations in $\mathbb{R}^n$, then the marginal distribution of the first $k$ coordinates $X_1,\cdots,X_k$ is close to a mixture of i.i.d. Gaussian distributions.  Here, the ``closeness'' is measured in the sense that the variation distance is bounded from above by $2(k+3)/(n-k-3)$ for $1 \leq k \leq n-3$.

Let us now formulate our main result, which is the quantum counterpart of the previous result.
\begin{theorem}
  If $\rho^{n}$ is a $n$-mode orthogonally invariant quantum state,
  its partial trace over any $(n-k)$ modes $\text{tr}_{n-k}(\rho^{n})$ can be   approximated in the sense of the trace-norm distance by a mixture of   $k$-mode thermal states $\rho_{\text{th}}^k(x)$, that is,
  \begin{eqnarray*}
    \lefteqn{ ||\text{tr}_{n-k}(\rho^{n})-\int \rho_{\text{th}}^k(x) \,       \mu(dx)||_1 } \hspace{1cm}\\
    && \leq 2 \left(\frac{n^2}{(n-k-1)(n-k-2)}-1 \right)
  \end{eqnarray*}
  where $\rho_{\text{th}}^k(x)$ is the tensor product of $k$ thermal states   with a mean number of $x$ photons per mode, and $\mu$ is a probability   measure.
\end{theorem}

The idea of our proof is inspired from that of the classical version of the theorem for geometric probability distributions, as described in \cite{DF87}. If $X_1,\cdots,X_n$ are integer classical random variables whose joint distribution is invariant under transformations that keep the sum $X_1+\cdots +X_n$ constant, then the marginal law of the first $k$ coordinates $X_1,\cdots,X_k$ is close, in the sense of the variation distance, to a mixture of i.i.d. geometric distributions. The link with our quantum problem comes from the fact that in the Fock basis, any passive linear interferometer redistributes the photons among the modes in such a way that the total photon number is kept constant, since the energy is conserved. The invariance under orthogonal transformations in phase space therefore translates into the invariance under transformations that keep the total photon number constant in the Fock basis. As a consequence, the asymptotic state in our theorem is characterized by a geometric distribution in the Fock basis, which precisely is the signature of a thermal state.  Our proof will thus consist in bounding the convergence of an $n$-mode state that is invariant under a redistribution of photons among the $n$ modes (with a constant total photon number) towards a mixture of thermal states.
  
\begin{proof}
  We start from the fact that any $n-$mode orthogonally invariant state   $\rho^{n}$ can be written as a convex mixture of the multimode number   states $\sigma_p^{n}$ as defined in Eq.~(\ref{defin}), namely
  \begin{equation}
    \rho^{n} = \sum_{p=0}^{\infty} c_p \, \sigma_p^{n}
  \end{equation}
  with arbitrary weights $c_p$ satisfying $0 \leq c_p \leq 1$ and $\sum_p c_p   =1$.  Now, using the convexity of the trace-norm distance
  \begin{align}
    & ||\text{tr}_{n-k}(\rho^{n})-\int \rho_{\text{th}}^k(x) \, \mu(dx)||_1     \nonumber \\
    & \leq \sum_{p=0}^{\infty} c_p \, ||\text{tr}_{n-k}(\sigma_p^{n})-\int     \rho_{\text{th}}^k(x) \, \mu(dx)||_1,
  \end{align}
  we see that it is sufficient to prove the theorem for the extremal states   $\sigma_p^{n}$, that is, it is enough to prove
  \begin{multline}
    ||\text{tr}_{n-k}(\sigma_p^n)- \rho_{\text{th}}^k(p/n)||_1 \leq \\
    2 \left(\frac{n^2}{(n-k-1)(n-k-2)}-1 \right),
    \label{eq-to-prove}
  \end{multline}
  for any $p$. Note that we have arbitrarily reduced the mixture of thermal   states to one single term, which is natural since we start with an extremal   state $\sigma_p^n$.  Note also that we have taken $x=p/n$ for this single   term, that is, we characterize the convergence of the reduced state towards   a $k$-mode thermal state with a mean number of $p/n$ photons per mode.

  The reduced state $\text{tr}_{n-k}(\sigma_p^n)$ is obviously orthogonally   invariant in the remaining space of $k$ modes, which implies that it can be   written as a mixture of $k$-mode number states,
  \begin{equation}
    \text{tr}_{n-k}(\sigma_p^n) = \sum_{l=0}^p f(l) \, \sigma_l^k 
  \end{equation}
  where a simple combinatorial argument shows that:
  \begin{equation}
    f(l)=\frac{a_l^k \; a_{p-l}^{n-k}}{a_p^n}.
  \end{equation}
  The $k$-mode thermal state $\rho_{\text{th}}^k(x)$ is defined as the   product of $k$ single-mode thermal states with $x$ photons per mode, namely   $\rho_{\text{th}}^k(x) = \rho_{\text{th}}(x)^{\otimes k}$ with
  \begin{equation}
    \rho_{\text{th}}(x)=\sum_{l=0}^{\infty} \frac{x^l}{(1+x)^{l+1}} \, \proj{l}
  \end{equation}
  A straightforward calculation shows that it can be written as a mixture of   $k$-mode number states
  \begin{equation}
    \rho_{\text{th}}^k(x) = \sum_{l=0}^{\infty} g(l) \, \sigma_l^k,
  \end{equation}
  with
  \begin{equation}
    g(l) =  a_l^k \frac{x^l}{(1+x)^{l+k}}
  \end{equation}
  which confirms that it is also orthogonally invariant.

  We now prove Eq.~(\ref{eq-to-prove}) using the fact that both   $\text{tr}_{n-k}(\sigma_p^n)$ and $\rho_{\text{th}}^k(x)$ are diagonal in   basis of $k$-mode number states. This implies that their trace-norm   distance is given by the variation distance between the classical   probability distributions $f$ and $g$, that is
  \begin{eqnarray}
    ||\text{tr}_{n-k}(\sigma_p^n)- \rho_{\text{th}}^k(p/n)||_1  &=&     \sum_{l=0}^{\infty}|f(l)-g(l)| \nonumber\\
    &=& 2\sum_{l=0}^{\infty}\left(\frac{f(l)}{g(l)}-1 \right)^+ g(l)     \nonumber \\
    &\leq& 2 \left(\sup_{l} \frac{f(l)}{g(l)} -1\right)
    \label{mainineq} 
  \end{eqnarray}
  where the function $(x)^+$ is equal to $x$ if $x\ge 0$ and vanishes   otherwise.  Using the notation
  \begin{equation}
    h(l) \equiv \frac{f(l)}{g(l)} = \frac{a_{p-l}^{n-k} \;       (1+p/n)^{l+k}}{a_p^n \;  (p/n)^l} ,
  \end{equation}
  the rest of the proof consists in upper bounding the supremum of $h(l)$ as   tightly as possible.  Expanding the binomials in $a_{p-l}^{n-k}$ and   $a_p^n$, the function $h(l)$ can be rewritten as:
  \begin{align}
    h(l)& = \frac{(n-1)!}{n^k \; (n-k-1)!} \times \frac{(p-1)!}{p^{l-1} \;       (p-l)!}  \\
    & \hspace{1cm} \times \frac{(n+p)^{k+l} \; (n+p-k-l-1)!}{(n+p-1)!}     \nonumber     \\
    & = \frac{\prod_{t=1}^k (1-\frac{t}{n})\prod_{t=1}^{l-1}       (1-\frac{t}{p})}{\prod_{t=1}^{k+l} (1-\frac{t}{n+p})}
  \end{align}
  The logarithm of $h(l)$ can be expressed as
  \begin{equation}
    \log h(l)= -S(n,k) - S(p,l-1)+S(n+p,k+l),
    \label{logh}
  \end{equation}
  where $S(n,k)$ is defined as
  \begin{equation}
    S(n,k) \equiv  -\sum_{t=0}^k \log \left(1-\frac{t}{n}\right).
  \end{equation}
  The function $x \mapsto -\log(1-x)$ being monotonically increasing on   $[0,1[$, one has
  \begin{equation}
    n \, J\left(\frac{k}{n}\right) \leq S(n,k) \leq n \,     J\left(\frac{k+1}{n}\right)
    \label{relationJ-S}
  \end{equation}
  where
  \begin{eqnarray}
    J(x) &\equiv& -\int_0^x \log(1-t)\, dt  \nonumber \\
    &=& x+(1-x) \log(1-x).  
  \end{eqnarray}
  Let us introduce the two reduced variables $u=k/n$ and $v=l/p$, which both   belong to the interval $[0,1[$.  Since the function $J(x)$ is convex on   $[0,1[$, we have
  \begin{equation}
    J(\alpha \, u +(1-\alpha)\,v) \le   \alpha J(u) + (1-\alpha)J(v)
  \end{equation}
  where $0\le\alpha\le 1$. If we choose $\alpha=n/(n+p)$, this equation   translates into
  \begin{equation}
    (n+p) \, J\left(\frac{k+l}{n+p}\right) \le
    n \, J\left(\frac{k}{n}\right) + p \, J\left(\frac{l}{p}\right) 
    \label{convexJ}  
  \end{equation}
  By using Eq.~(\ref{relationJ-S}), we can lower (upper) bound the left-   (right-) hand side term of Eq. (\ref{convexJ}), which yields
  \begin{equation}
    S(n+p,k+l-1) \le S(n,k) + S(p,l)
  \end{equation}
  Substituting $k$ with $k+2$ and $l$ with $l-1$, we get the equivalent   inequality
  \begin{equation}
    S(n+p,k+l) \le S(n,k+2) + S(p,l-1)
  \end{equation}
  which can be used to upper bound the quantity of interest obtained in Eq.   (\ref{logh}), namely
  \begin{equation}
    \log h(l) \le S(n,k+2)-S(n,k)
  \end{equation}
  We conclude that
  \begin{equation}
    h(l) \le \frac{n^2}{(n-k-1)(n-k-2)}
  \end{equation}
  which, using Eq.~(\ref{mainineq}), concludes the proof of our theorem.

\end{proof}

\section{Conclusion}
\label{conclusion}

We have investigated a new type of symmetry in the context of quantum de Finetti's theorems, namely the invariance under orthogonal transformations in phase space representation. This approach seems to be particularly relevant to study the properties of continuous-variable systems, going beyond the standard approach that was based on permutation invariance in state space representation. Just like orthogonally invariant $n$-partite probability distributions are known to tend to i.i.d. Gaussian distributions, we have shown that orthogonally invariant $n$-mode states tend to i.i.d. thermal states.  More precisely, we have derived a finite version of a quantum de Finetti's theorem for this class of states, which puts an upper bound on the distance between the partial trace of orthogonally invariant states and mixtures of multimode thermal states. Physically, the invariance under orthogonal transformations in phase space corresponds to the fact that the state is unchanged by a passive linear interferometer. Since this operation amounts to redistributing photons while keeping their number constant, our quantum de Finetti's theorem is connected to the classical de Finetti's theorem for geometric distributions where geometric distributions (thermal states) play a special role.

Let us conclude by suggesting two potentially interesting extensions of this de Finetti theorem, which arise in the context of continuous-variable quantum key distribution \cite{CG07}.  First, it would be nice to generalize our results to bipartite states, i.e., states $\rho_{AB}$ that are invariant under (conjugate) orthogonal transformations applied to systems $A$ and $B$, respectively.
As we explained in Ref.~\cite{LKGC09}, the legitimate parties (Alice and Bob) can always enforce such a symmetry in phase space. Their global state $\rho_{AB}$ can therefore be assumed to be a bipartite orthogonally invariant state in phase space. In other words, $\rho_{AB}$ is invariant if both parts $\rho_A = \mathrm{tr}_B \; \rho_{AB}$ and $\rho_B = \mathrm{tr}_A \; \rho_{AB}$ are processed via (conjugate) passive linear interferometers. Note that the resulting local states held by each party, $\rho_A$ and $\rho_B$, are then another example of orthogonally invariant states.  The second question one might want to answer is whether the de Finetti theorem presented here has an exponential version in analogy to Ref.~\cite{R07}, that is, such that only a small number of modes needs to be traced out in order to get a reduced state that is well approximated by (almost) a mixture of thermal states.

\section*{Ackowledgements}
A.L. thanks Renato Renner and Johan {\AA}berg for fruitful discussions.  The authors acknowledge financial support of the European Union under the FET-Open project COMPAS (212008), of Agence Nationale de la Recherche under projects PROSPIQ (ANR-06-NANO-041-05) and SEQURE (ANR-07-SESU-011-01), and of the Brussels-Capital Region under project CRYPTASC.


\end{document}